\documentclass[10pt,a4paper]{article}

\usepackage[a4paper,margin=1in]{geometry}
\usepackage{oejformat}
\usepackage{authblk}
\usepackage{graphicx}
\usepackage{mwe}          
\usepackage{booktabs}
\usepackage{amsmath}
\usepackage[hidelinks]{hyperref}
\usepackage{enumitem}
\usepackage{caption}
\usepackage{textgreek}
\usepackage[numbers,sort&compress]{natbib}

\title{Topological beam stirring in a multicore fiber}
\author[1,2\corrauthor]{A.Y. Kolesnikova}
\author[1,2]{M.D. Gervaziev}
\author[1,2]{N.V. Bochkarev}
\author[1,2]{D.S. Kharenko}
\author[1,2]{E.V.~Podivilov}
\author[1,2\corrauthor\corrauthor]{S.A. Babin}
\affil[1]{Insitute of Automation and Electrometry SB RAS, Novosibirsk 630090, Russia}
\affil[2]{Novosibirsk State University, Novosibirsk 630090, Russia}

\date{} 

\begin{document}

\maketitle

\begin{abstract}
Multicore fibers (MCF) are perspective media for telecommunications, sensing, imaging and laser technologies. Here, the effect of beam stirring between weakly coupled cores is observed for sub-nanosecond transform-limited pulses of several kW peak power propagating in \(\sim \)10 m long 7-core fiber, for the first time to our knowledge. In contrast to low-power domain where the output power distribution in the cores is random with large fluctuations sensitive to fiber disturbances, at high power of the input pulse injected in the central core the output power becomes equalized between the cores with fluctuations reduced to \(<5\%\) being insensitive to disturbances. Similar behavior is observed in cut-back experiments showing that equi-partition is approached at a distance of \({\sim} 5\)~m. The performed modeling describes well the experimental results and clarifies mechanisms of the new effect reasoned by a large nonlinear phase shift changing along the pulse and thereby resulting in statistical averaging over the pulse length of multidirectional power transfer processes between cores, thus leading to the robust equilibrium (equi-partitition for hexagonal MCF topology). At the same time, the combined output beam measured in a far field takes a stable bell-shaped profile instead of speckled beam at low powers, similar to the beam self-cleaning effect in multimode fibers.
\end{abstract}

\keywords{multicore fiber, Kerr nonlinearity, beam stirring}

\section{Introduction}
Multicore fibers (MCFs) were initially proposed and implemented as one of the possible methods of spatial division multiplexing (SDM) for high-capacity optical communications, which may overcome the limitations of a conventional singlemode fiber (SMF)\,\cite{b1}. The MCF-based SDM technology is treated as much more cost- and power-efficient in comparison with parallel deployment of multiple SMFs and may be less affected by a cross-talk between parallel spatial channels in comparison with multimode fibers where significant overlap between transverse modes is always present. The cross-talk in MCFs may be greatly reduced by increasing the spacing between the cores. The potential of the MCF-based SDM technology has been recently confirmed by experimental demonstration of >10 Pb/s transmission\,\cite{b2}.

Another application of MCFs with uncoupled cores is their use for distributed sensing also based on the unique capability of independent light transmission in multiple spatial channels. In comparison with standard SMF, the MCF bending gives rise to tangential strain in off-center cores which may be employed for directional bending and 3D shape sensing\,\cite{b3}. On the other hand, the parallel cores enable SDM system configuration that allows for the multiplexing of multiple distributed sensing techniques such as optical frequency-domain reflectometry (OFDR), Brillouin optical time domain analysis (BOTDA) or fiber Bragg gratings (FBG) interrogation. As a result, multi-parameter sensing can be achieved by using single MCF. Besides, MCFs are prospective for imaging\,\cite{b4}.

Implementation of FBG arrays inscribed in MCFs by femtosecond laser pulses with high spatial resolution over transverse and longitudinal coordinates \cite{b5} offers new opportunities in 3D shape sensing, as well as in new designs of fiber laser cavities. The selective inscription of FBGs with individual parameters in different cores allows to control spatio-spectral characteristics of output beam in CW fiber lasers based on MCFs. The individual cores may generate either independently delivering multiple uncoupled beams at different wavelengths defined by individual FBGs or generate strongly coupled beams in coupled-core MCF with the output localized in one core \cite{b5,b6} and/or at single wavelength corresponding to the maximum of geometric-mean reflection of individual FBGs \cite{b7}. At the same time, the closely spaced coupled cores may have large mode area (LMA) that offers power enhancement at the intensity kept at rather low level thus limiting nonlinear effects such as Kerr-nonlinearity induced spectral broadening \cite{b8}.   

MCFs are also treated as one of the perspective approaches for power scaling of short-pulse laser systems \cite{b9} being a more reliable alternative to a rather complicated and expensive spatially separated fiber amplifiers with coherently combined output which already outperform single-emitter systems \cite{b10}. The implementation of active MCFs using a single-beam seed divided among independent amplifying LMA cores and a subsequent combining of output beams with active control of their phases enables a kW-level average power of ultrashort pulse trains in coherently combined systems \cite{b9}. At the same time, in coupled-core MCFs close arrangement of multiple doped cores in a shared cladding leading to evanescent-field coupling of the guided waves results in either internal self-organization or external formation of supermodes (with locked phases in individual cores) being kept unchanged over long fiber lengths and resistant to nonlinear effects. In this way, an efficient out-of-phase (with \(\pi\)-phase shift in neighbor cores) supermode amplification of short pulses in a circular 6-core Yb-doped fiber \cite{b11} or in a MCF with square N\(\times\)N array of coupled cores \cite{b12} is shown to be resistant to discrete modulation instability and can carry total power greatly exceeding the self-focusing threshold for single-core fiber. 

The nonlinear effect of self-focusing manifests itself in MCFs as a light localization within a few (or even single) cores accompanied by a pulse length compression at increasing energy \cite{b13}. The numerical simulation predicts nonlinear combining into a single core of \(\sim90\% \) of the total energy of pulses initially injected into all cores of a hexagonal 7-core MCF and a pulse compression factor of about 720 in a 19-core ring MCF. Herewith, the self-focusing based nonlinear combining is not sensitive to relative phase mismatches. Such pulse localization/compression in the central core of a hexagonal 7-core fiber was experimentally observed for sub-\(\mu\)J 370 fs pulses compressed to 53 fs length with corresponding peak power of >2 MW \cite{b14}. 

Here we numerically and experimentally demonstrate, for the first time to our knowledge, another nonlinear effect in a hexagonal 7-core MCF as an example, which manifests itself at pulse propagation with much lower (kW) peak powers, far from the self-focusing threshold. In the presence of random optical coupling of the cores (e.g. at fiber coiling) the pulse is randomly redistributed between the cores regardless of the initial conditions (e.g. injected in a single core) with 100\% power fluctuations in individual cores as the pulse propagates along the fiber. It is revealed that at kW-level peak power, the fluctuations greatly reduce at the pulse propagation over tens meters of MCF so that the power in the individual cores is exactly equalized and stabilized becoming resistant to external disturbances of the fiber, in contrast to the case of linear random coupling. The dependence of the nonlinear stabilization length on the pulse peak power and the coupling strength of MCF cores, as well as the amplitude and characteristic length of the refractive index fluctuations in the cores is studied numerically and compared with experiment. Based on the results of the numerical and experimental investigations, qualitative explanation of this new effect is provided.

\section{Experiment}
\begin{figure}[htbp!]
    \centering
    \includegraphics[width=\linewidth]{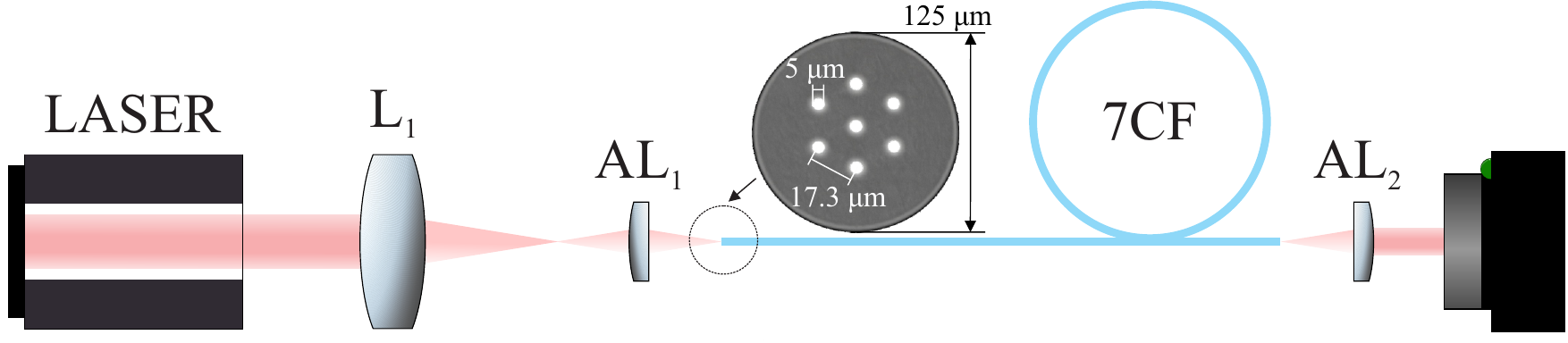}
    \caption{Experimental setup: \(\text{L}_1\) – lens, \(\text{AL}_{1,2}\) – aspheric lenses for beam collimation at the output of the 7CF, the image of the fiber output beam is captured with CMOS camera}
    \label{fig:1}
\end{figure}

In our experiments we launched the beam from narrowband q-switched microchip laser Standa STA-01-7-OEM (central wavelength \(\lambda = 1064\,\)nm, pulse duration 435 ps, repetition rate 1 kHz) into the passive 7-core fiber (7CF) produced by FORC RAS (see \cite{b5} for details). A schematic of the setup is demonstrated on Fig. \ref{fig:1}. In this system, the transmission reached \(55\%\) while the laser beam is optimally focused in a core. The output power distribution between the cores was recorded using a Gentec Beamage-4M beam profiler. To minimize/average the influence on the studied spatial dynamics of refractive index fluctuations along a core, an 11 m long piece of fiber was coiled into a ring with a diameter of 6 cm.

\begin{figure}[htbp!]
    \centering
    \includegraphics[width=0.5\linewidth]{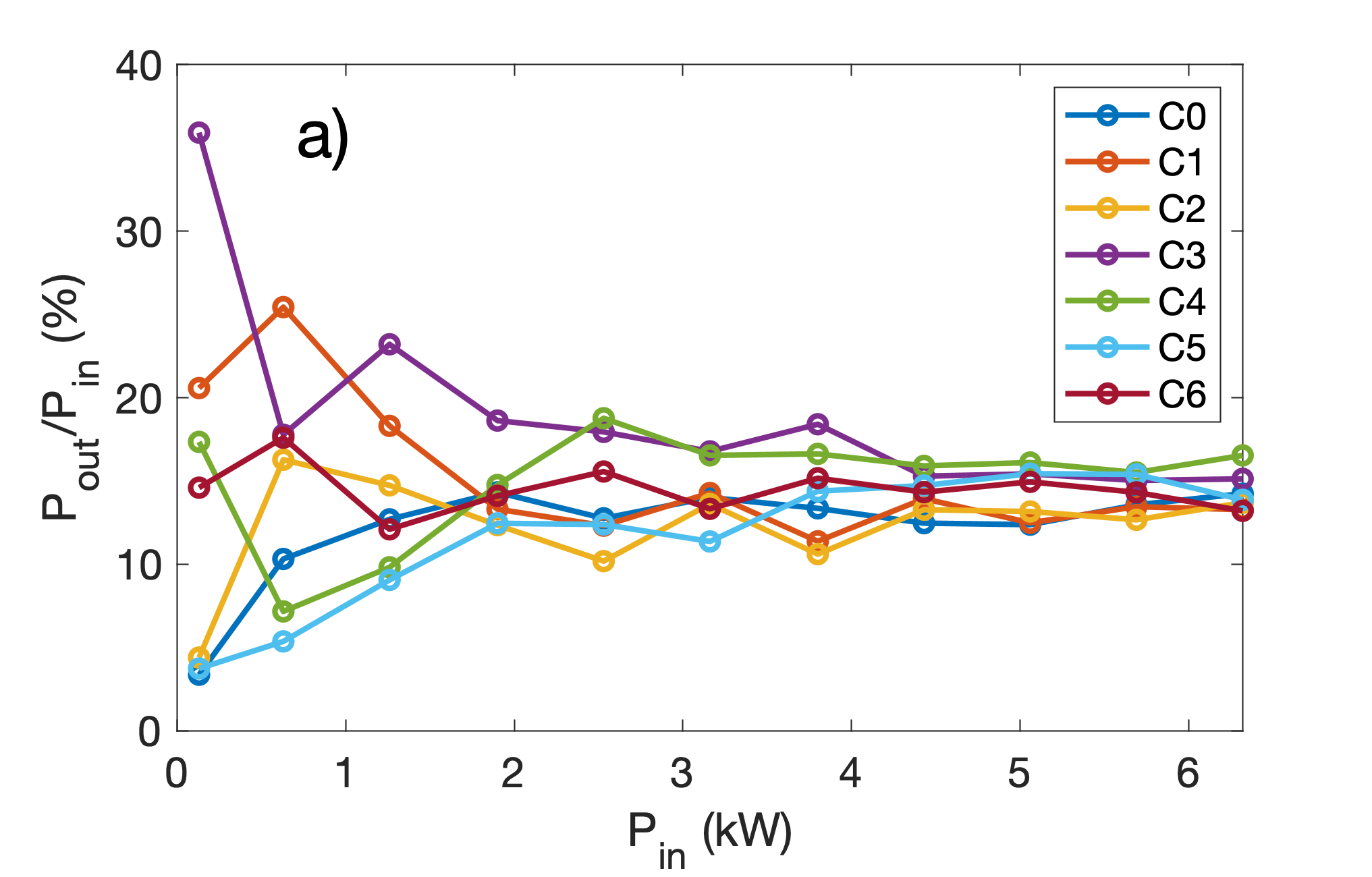}\includegraphics[width=0.5\linewidth]{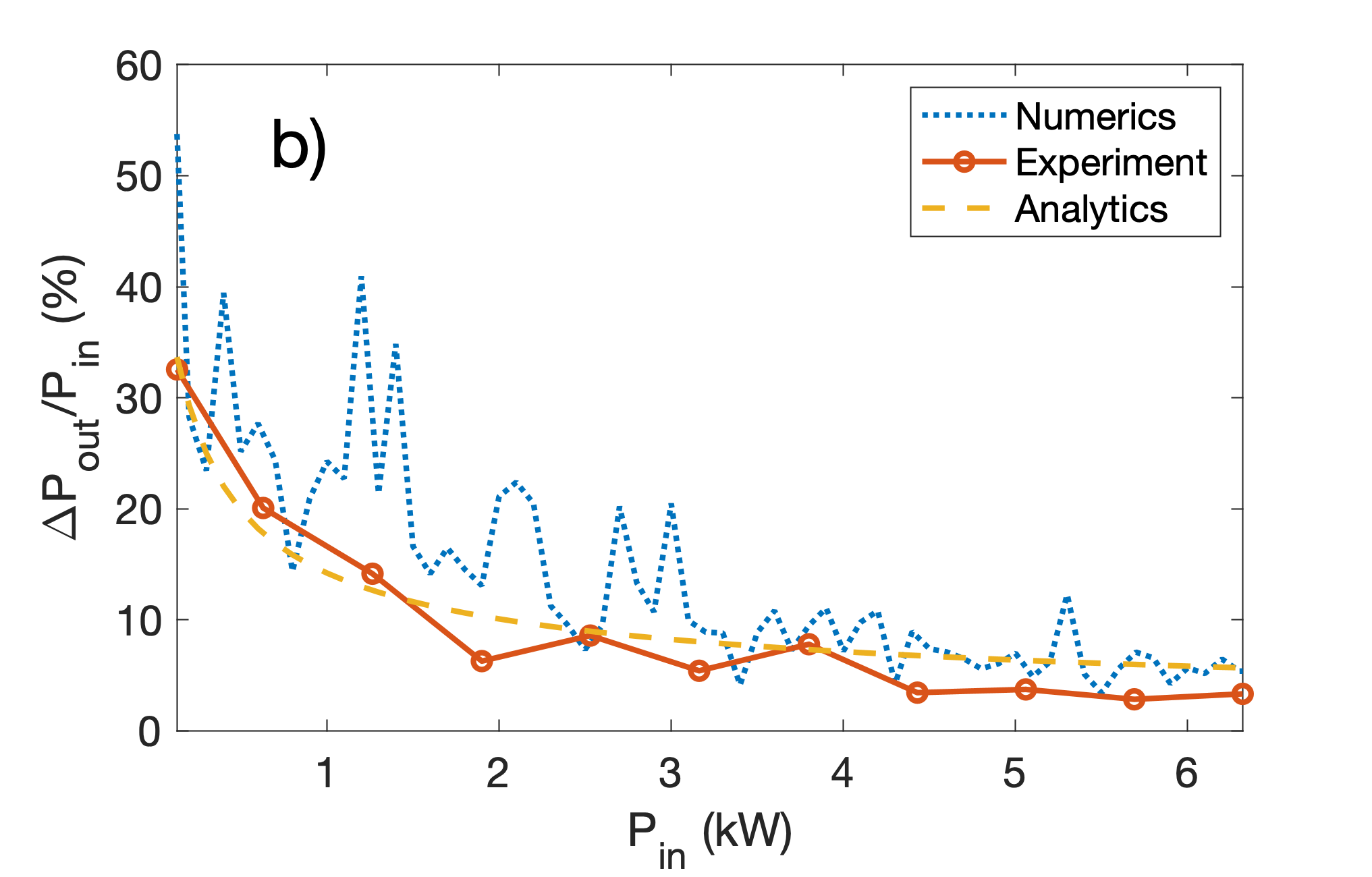}
    \caption{a) The output power fraction in individual cores (C0-C6, C0 is central) as a function of the input peak power of the pulse propagating in a 11-m 7CF coiled into a ring with diameter of 6 cm; b) the relative difference between the maximum and minimum output power (core fractions) as a function of the input peak power (blue line corresponds to experimental data, orange line – to numerical modelling, yellow line – to analytical approximation). }
    \label{fig:2}
\end{figure}

The obtained experimental data were used to plot the output power distribution of the each cores C0-C6 as a function of the input pulse peak power (\(P_\text{in}\)), taking the transmission into account (Fig.\,\ref{fig:2}a). At relatively low power (\(P_\text{in} = 0.13\,\)kW) the beam initially injected in the central core randomly redistributed between all the cores. As we increase the input power the observed power deviations in the cores start to reduce reaching an almost complete equalization (with fraction value of \(\sim 14\%\) for all the cores) at \(P_\text{in} = 6.32\,\)kW. We did not observe significant changes in the input/output power transmission ratio between the low- and the high-power excitation regimes, which indicates that the effect does not arise due to a change in the linear losses. To demonstrate the nonlinear nature of the equalization effect, we plotted the difference between the maximum and minimum output power fraction in the cores as a function of the input peak power (Fig.\,\ref{fig:2}b). As we increased the input peak power the measured spread of the output power fractions dropped down from \(>30\%\) to \(<5\%\). This key experimental dependence illustrating the new nonlinear effect of spatial beam stirring in MCFs is used for further comparison with the results of analytical and numerical modeling described below.

By measuring optical spectra with the use of Yokogawa AQ6370D analyzer we determined that the spatial beam stirring occurs with the narrowband optical spectra kept almost unchanged within the resolution of the spectrum analyzer. It should be noted that at the maximum power a shifted Raman scattering peak becomes visible, but the equi-distribution is already formed at lower powers.

\begin{figure}[htbp!]
    \centering
    \includegraphics[width=0.7\linewidth]{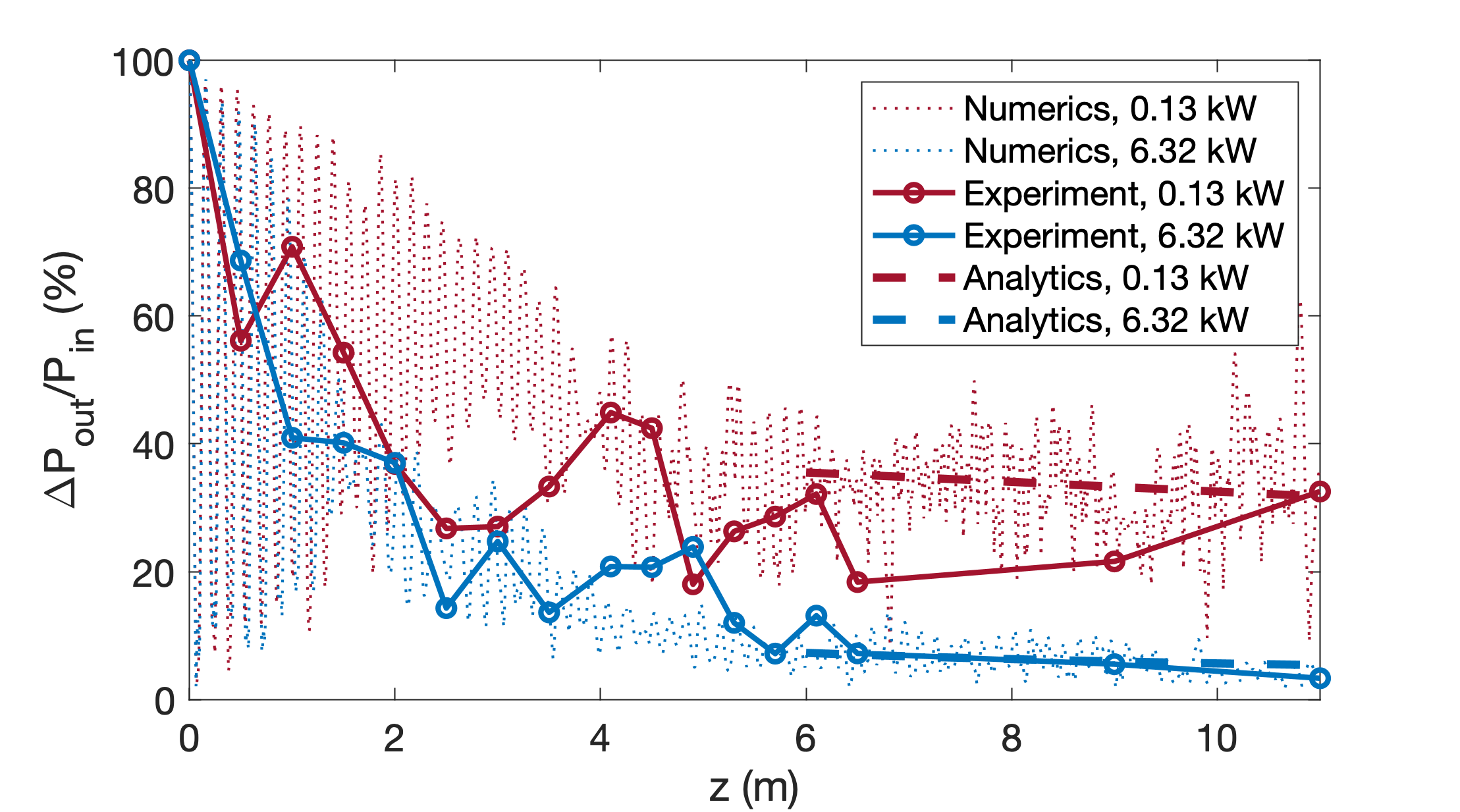}
    \caption{The relative difference between the maximum and minimum power (core fractions) versus fiber length for three input peak power values. }
    \label{fig:3}
\end{figure}

For deeper insight into the nonlinear dynamics of the beam propagating in 7CF with randomly coupled cores, we performed a study of its evolution along the fiber by its gradually backward cutting at the injection conditions unchanged. The relative power difference in the cores starting from the input facet (corresponding to the launch into the central core) up to 11 meters length is demonstrated in Fig.\,\ref{fig:3} for two different levels of \(P_\text{in}\). Over the first two meters, the beam undergoes a similar evolution, showing a decrease in the spread between the core fractions. However, the dynamics diverge significantly at longer distances. At low power the spread between the core fraction difference fluctuates around \(30\%\) and does not reach equalized powers. For higher input powers, the difference continues decreasing. At that the equi-partition emerges at different length depending on power so that at the maximum length of 11 m the difference decreases with increasing power as shown in Fig.\,\ref{fig:2}b. 

\begin{figure}[htbp!]
    \centering
    \includegraphics[width=0.85\linewidth]{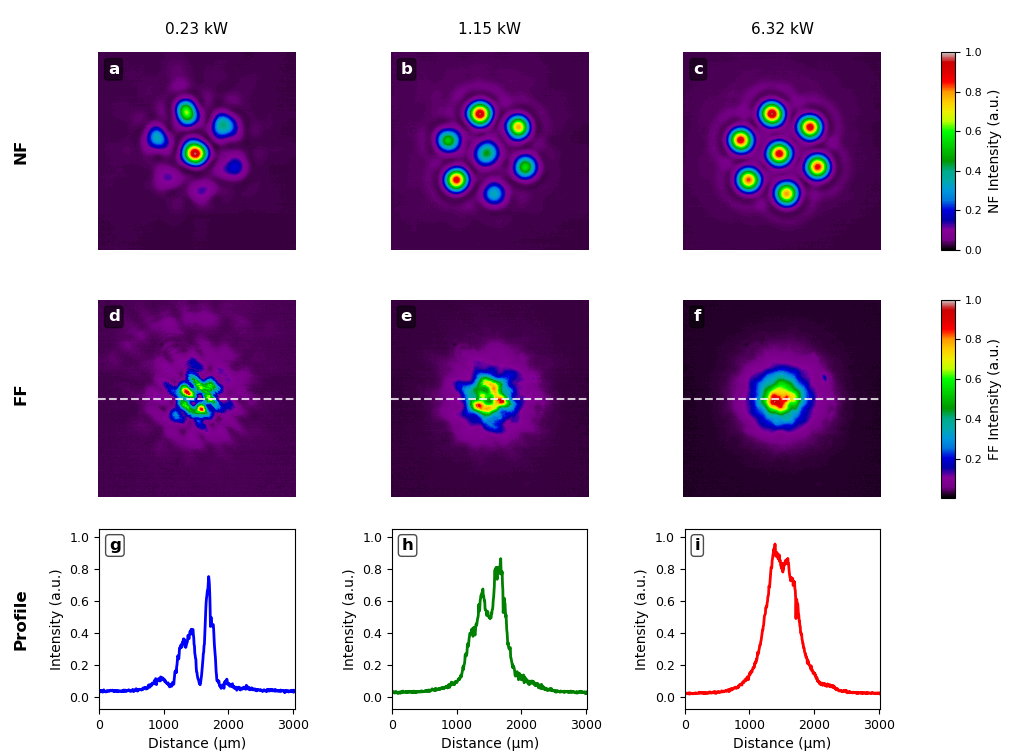}
    \caption{NF images of the 7CF output captured at 1064 nm (a-c) demonstrate the spatial beam stirring, namely the formation of 7 beams with equal intensities when the \(P_\text{in}\) increases. FF images (d-f) demonstrate the coherent combining of the beams with the establishment of bell-shaped transverse profile (g-i). Fiber length - 11 m.}
    \label{fig:4}
\end{figure}

Finally, we studied the evolution of transverse beam profiles at the 11-m MCF output. We captured the near field (NF) and – using the Fourier lens – far field (FF) images (Fig.\,\ref{fig:4}). At low power (Fig.\,\ref{fig:4}a), the NF power distribution between cores is random, with the maximum in one of the peripheral cores, whereas some other ones almost have no power in them. At high power (Fig.\,\ref{fig:4}c) equi-partition was established, and all the cores have nearly equal output power. At the same time, FF images demonstrate the result of coherent combining of the beams corresponding to the MCF cores. It is noteworthy that at low power (Fig.\,\ref{fig:4}d), such a beam has a speckled structure, but at high power the beam acquires a characteristic bell-shaped profile with enhanced brightness and quality (Fig.\,\ref{fig:4}f). This dynamic resembles another notable spatial nonlinear effect, namely spatial Kerr self-cleaning, which occurs in multimode fibers \cite{b15}. Additionally, we investigated the stability of the output spatial dynamics against external localized mechanical stress. At low powers, external disturbances applied to the fiber significantly altered the FF speckle structure, as well as the NF intensity distribution between the cores~(see the videofiles "NF$\_$Low$\_$Power" and "FF$\_$Low$\_$Power" in the supplementary material). So, the beam stirring demonstrated is a robust effect being insensitive to disturbances such as bending and twisting~(see the videofiles "NF$\_$High$\_$Power" and "FF$\_$High$\_$Power" in the supplementary material) which reinforces the similarity with Kerr beam self-cleaning. This analogy requires a deeper understanding of key physical mechanisms of the  effect that may be achieved by the development of the appropriate model.

\section{Modelling and comparison with experiment}

The pulse propagation model in a multicore fiber can be written as follows. The electric field is the sum of the pulses propagating in the cores \(E=\sum_{j=0}^6A_j(z,t)\Psi(\mathbf{r-r}_j ) e^{i\omega t-ikz}\) with the carrier frequency \(\omega=281\) THz (\(\lambda\)=1064 nm) and the propagation constant \(k=(2\pi n_\text{eff})/\lambda=8.6\)\,\textmu m\textsuperscript{-1}, where \(n_\text{eff}\) is the average effective refractive index in the cores. Here \(\Psi(\mathbf{r-r}_j )\)  is the normalized spatial distribution of core \(j\) fundamental mode in fiber cross-section, \(r_j\) is the coordinate of core \(j\) center (\(j = 0,..,6\)). The strength of the coupling between the modes of different cores is determined by the overlap integral \(J\sim\int(n_2-n_1)/n_1\Psi(\mathbf{r-r}_j )\Psi(\mathbf{r-r}_k ) d^2 r  (\forall j\neq k)\), and its propagation evolution is described by a system of coupled equations:

\begin{equation}
    \begin{cases}
    \cfrac{dA_0}{dz}=i\delta n_0(z) k A_0+
    i\left(k+\cfrac{in_\text{eff}}{c}\cfrac{d}{dt}\right)J\sum_{j=1}^6A_j+i\gamma|A_0|^2A_0\\
    \cfrac{dA_j}{dz}=i\delta n_j(z) k A_j+
    i\left(k+\cfrac{in_\text{eff}}{c}\cfrac{d}{dt}\right)J\left(A_0+A_{j-1}+A_{j+1}\right)+i\gamma|A_j|^2A_j
      \end{cases}
    \label{Eq_1}
\end{equation}
Here the square of the field amplitude is normalized to the power. The coefficient of self-phase modulation due to Kerr nonlinearity is \(\gamma=5\cdot10^{-3} \text{(W}\cdot \text{m})^{-1}\) in each core. The addition of a time derivative in the linear coupling term is necessary to account the different group velocities of the supermodes. The model includes the relative difference in refractive indices \(\delta n_j (z)\) in the cores arising at fiber fabrication and its winding on a spool. It is randomly varied such that \(\sum_{j=0}^6 \delta n_j=0\) and does not exceed a certain specified value \(|\delta n_j |<\delta n_\text{max}\). The initial state is the pulse injected to the central core at the input \(A_0(0,t)=\sqrt{P_\text{in}}/\text{cosh}(t/\tau)\), where \(\tau\) is a pulse duration, \(P_\text{in}\) is a peak input power.

The numerical scheme for solving the system of equations \eqref{Eq_1} is based on the Split-Step method, where the linear and nonlinear parts of the equation are calculated independently; the linear part is calculated using the matrix exponent. To calculate the time derivative in a step with a linear coupling of cores, the Fourier transform in time was used. The time step is \(\tau=5\,\)ps, the spatial step is \(h=250\)\,\textmu m. The relative difference in refractive indices \(\delta n_j\) changes with a characteristic length \(L_\text{ref}=1\) mm, the intermediate values \(\delta n_j (z)\) are determined by a spline.

In order to describe the experimentally observed beam stirring, the parameters of the coupling strength J and the maximum variation of the refractive index \(\delta n_\text{max}\) were varied. First of all, the simulation results showed that the beam stirring among the cores is observed only in the case of relatively weak coupling \(J<\delta n_\text{max}\), while for strong coupling a non-trivial intensity distribution across the supermodes is observed. For each experimental power, about 150 simulations were carried out with parameters varying in the reasonable range \(\delta n_\text{max}\in[10^{-6},10^{-4}]\) and \(J\in[10^{-7},10^{-5}]\). The most suitable parameters for describing the experimental results were chosen (\(\delta n_\text{max}=3\cdot 10^{-6}\) and \(J=9\cdot 10^{-7}\) for which the numerical simulation agrees well with the experimental data for output power difference in the cores as a function of  input peak power (see Fig. \ref{fig:2}b). The main difference of the simulated curve is the presence of short-scale fluctuations which are not pronounced in the experimental data consisting of separated points. 

\begin{figure}[htbp!]
    \centering
    \includegraphics[width=0.8\linewidth]{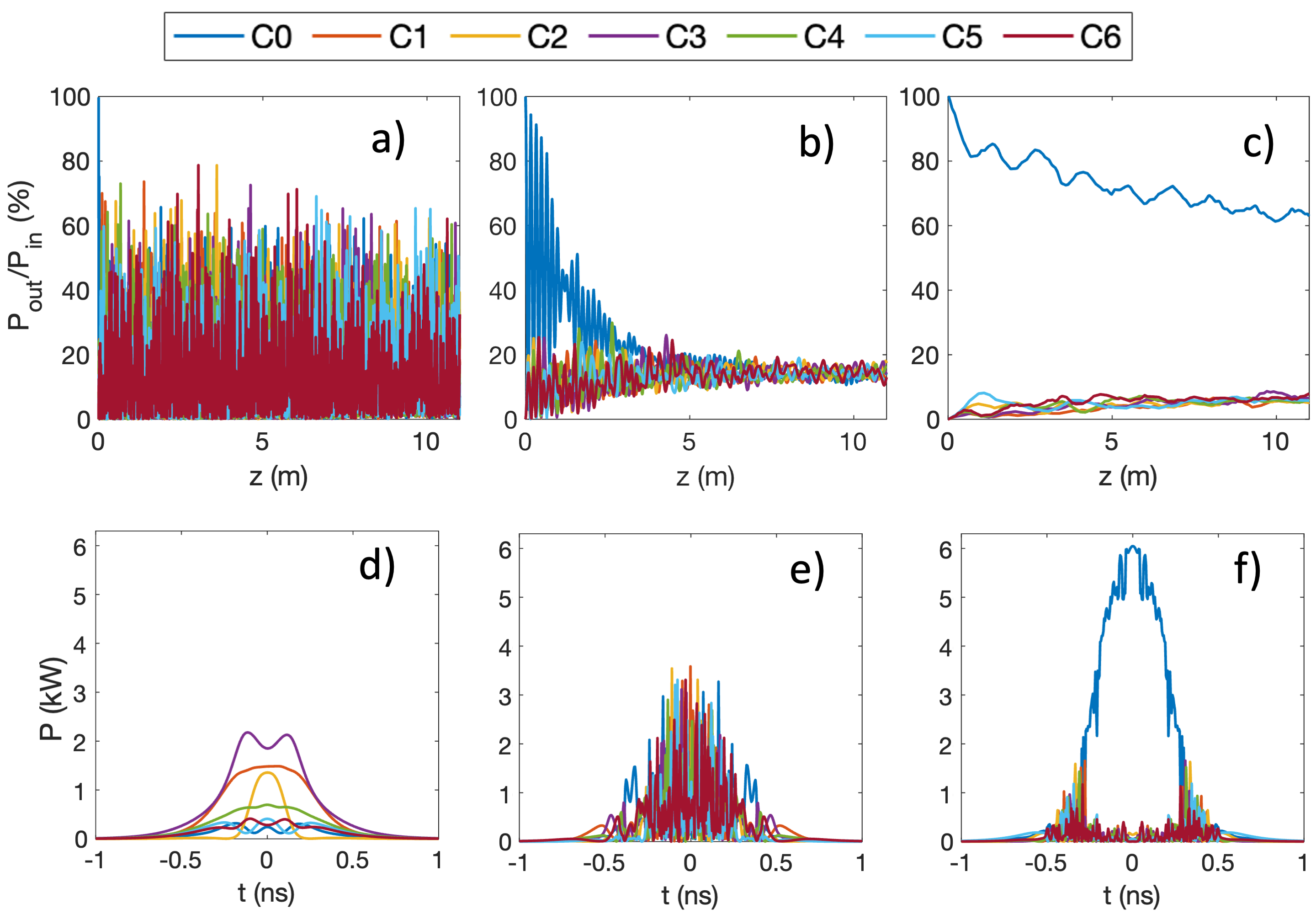}
    \caption{The power distribution along the fiber in different cores normalized to input power \(P_\text{in} =6.32\,\)kW and corresponding output pulses at MCF parameters: a, d) \(\delta n_\text{max}=3\cdot 10^{-5}\), \(J=9\cdot10^{-6}\); b, e) \(\delta n_\text{max}=3\cdot 10^{-6}\), \(J=9\cdot10^{-7}\), c, f) \(\delta n_\text{max}=3\cdot 10^{-6}\), \(J=10^{-7}\). }
    \label{fig:5}
\end{figure}

Simulation with the same parameters of the longitudinal power variations in the individual cores at maximum input power \(P_\text{in} =6.32\) kW is shown Fig.\ref{fig:5}b. The input power injected in the central core starts oscillating between the fiber cores with \(\sim 100\%\) relative variations at the initial part of the MCF. After 3--5 m the oscillations amplitude is greatly reduced whereas the average powers tend to equi-partition reached at \(\sim 10\)m \(\sim 14\%\) in each core with fluctuations of \(<5\%\)) that agrees well with the experimental data (Fig.\ref{fig:3}). It's important to note that the power equi-partition in the cores occurs not simply in the case of weak coupling (\(\delta n_\text{max}/J=3.3\) for the best-fit data), but also depends on absolute values. For example, if we increase the coupling strength \(J\) and refractive index variation \(\delta n_\text{max}\) by an order of magnitude, we observed chaotic power oscillations in the cores with no change in their amplitudes with distance along the fiber (see Fig.\,\ref{fig:5}a). In this case, the shape of the pulse in the cores is slightly deformed, since the nonlinear phase is disrupted due to a large change in the refractive index resulting in predominantly linear (randomly varying) redistribution of the whole pulse (see Fig.\,\ref{fig:5}d). Besides, a decrease in coupling strength with experimental refractive index variation prevents linear interaction between the cores, and the equi-partition effect disappears (Figs. \ref{fig:5}c,\,f) with the pulse predominantly propagating in the central core.

In the case of optimal values of the refractive index variation \(\delta n_\text{max}\) and the coupling strength \(J\), the effective length of energy transfer between the cores at high peak power turns out to be greater than the nonlinear length \(L_\text{NL}=1/(\gamma P_\text{in}/7)\). A large nonlinear phase shift leads to the emergence of oscillations along the pulse (see Fig.\,\ref{fig:5}e), and their number \(N=\sqrt{1+(\gamma P_\text{in}z/7)^2}\) is determined by the nonlinear phase incursion in each core in average. So, after averaging over pulse length the power fluctuations in the cores are reduced by a factor of \(\sqrt{N}\), and their magnitude depends on the input power and the distance according to the following law:
\begin{equation}
    \cfrac{\Delta P(z,P_\text{in})}{P_\text{in}}=\cfrac{\Delta P(z=0)}{P_\text{in}}\left(1+((\gamma P_\text{in}z)/7)^2\right)^{-1/4}
\end{equation}

This analytical approach is in good agreement with the experimental and numerical data for the dependence on both the pulse peak power (Fig.\,\ref{fig:2}b) and on the propagation distance (Fig.\,\ref{fig:3}) taking into account that it is applicable for relatively long distances. 

In addition to the equi-partition between the cores, the statistical averaging over \(N\) sub-pulses may also be a reason of the bell-shaped beam formation at high peak power (Fig.\,\ref{fig:4}c,\,f,\,i), in contrast to the speckled structure observed at low power (Fig.\,\ref{fig:4}a,\,d,\,g) when the whole pulse is randomly distributed between the cores with the subsequent interference of partial beams in far field. 

\section{Results and discussion}
So, the effect of beam stirring between the weakly coupled cores of multicore fiber experimentally observed for sub-nanosecond transform-limited pulses of several kW peak power propagating in 11-m long 7-core fiber appears to be a new nonlinear effect. In contrast to low-power domain where the output power distribution in the cores is random with large fluctuations sensitive to fiber disturbances, at high power of the input pulse injected in the central (or another) core the output power becomes equalized between the cores with fluctuations reduced to \(<5\%\) being insensitive to disturbances. Similar behavior is observed in cut-back experiments showing that at short propagation distances even high-power pulse is randomly distributed between the cores approaching equi-partition with increasing distance. At the same time, the output beam with equalized core powers measured in a far field tends to a stable bell-shaped profile, in contrast to strongly speckled unstable beam observed at low powers.      

Based on the performed numerical modeling and analytical treatment, the observed nonlinear beam stirring effect has a statistical nature. At low peak power, a weak optical coupling of the core modes in the presence of random refractive index variations results in random redistribution between the cores of the pulse as a whole, with fluctuations of power difference in the cores of the order of their average values. At high peak power, the nonlinear phase shift with increasing propagation distance becomes large (being inconstant along the pulse) thus changing the direction of power conversion between the cores. As a result, the power starts oscillating along the pulse in all cores, the number of periods (sub-pulses) reaches \(N\sim50\) in our experiment. Though within each \(\sim 10\)-ps sub-pulse the power distribution between the cores is random, averaging in time domain over the whole pulse results in reduction of fluctuations by a factor of \(\sqrt{N}\) , so the power distribution is equalized and stabilized. Similar averaging of different speckled patterns in spatial domain arising from interference of partial beams within the sub-pulses leads to the formation of a bell-shaped stable beam in a far field.  

In its external manifestations, the beam stirring effect in MCF looks quite similar to the spatial beam self-cleaning (BSC) observed earlier in multimode fiber (MMF) at nearly the same pulse parameters and fiber length \cite{b15}. As it was shown later \cite{b16,b17}, the BSC effect may be described by statistical mechanics. At low peak powers of a pulse propagating in a graded-index MMF (experiencing random mode coupling due to bending etc.) the distribution of transverse modes tends to random one with nearly equal average values resulting in a highly speckled output beam. At kW-level peak power when the nonlinear (Kerr) four-wave mixing of modes comes into play, the output beam takes a smooth bell-like shape with mode distribution over their quantum number approaching the Rayleigh-Jeans law (RJ) with decreasing number of higher orders \cite{b17} that corresponds to  thermodynamic equilibrium.  In case of MCF, instead of transverse modes we have different core modes and their equi-partition instead of RJ distribution, but with greatly stabilized core powers and reduced fluctuations. The robustness of the beam stirring effect indicates that a sort of equilibrium distribution is established in MCF with weakly coupled cores. It may be defined by nearly equal topological properties of the cores in hexagonal MCF so that this equi-partition is a statistical result of thermodynamic equilibrium and the source of statistics is clear – large nonlinear phase shift leading to formation of large number of sub-pulses which are statistically averaged along the pulse. In this case the statistical approach is applicable to a single pulse (as well as to pulse trains). Though the reason of statistical behaviour is not discussed in detail for the BSC effect in MMF (see \cite{b16,b17} and citation therein), large nonlinear phase shift may play a similar role there.

\section{Conclusions}
Thus, the novel effect of beam stirring is quite important for the fundamental research and development of nonlinear photonics in multimode and multicore fibers, paving the way for new technologies and their practical applications. In particular, it may be applied for power-equalized light amplification in the weakly coupled cores of active MCFs to extremely high powers. The effect of beam quality improvement in far field is also interesting for applications such as high-power beam combining and nonlinear imaging, but it requires further exploring the details, namely phase properties of the partial beams and combined beam, which define coherent or incoherent regime.  For the sake of knowledge completeness and future developments it is also interesting to investigate the effect for MCFs of various topologies and high-peak-power pulses of various length and spectrum.

\section*{Acknowledgements}
The work is supported by Russian Science Foundation (grant 21-72-30024-Π) and the work of A.Yu.Kolesnikova on simulation – by Ministry of Science and Higher Education of the Russian Federation (FSUS-2025-0010).

\section*{Author contributions}
A.Y.K. made numerical model and performed the simulation. M.D.G. and N.B. assembled the experimental setup and performed the measurements. E.V.P. derived the model and make analytical estimations. D.S.K. supervised the experimental part of the project. S.A.B. proposed the original idea and supervised the project. All the authors wrote the paper. A.Y.K. and M.D.G. contributed equally to this work.

\section*{Competing interests}
The authors declare no competing financial interests.

\section*{Supplementary information}
Supplementary information for this paper is available at https://doi.org/10.29026/xxx.20xx.xxxxxx.

\section*{Authors}
*Correspondence to: A.Y. Kolesnikova, Novosibirsk State University, Novosibirsk 630090, Russia. Email: a.kolesnikova@g.nsu.ru. **S.A. Babin, Institute of Automation and Electrometry SB RAS, Novosibirsk 630090, Russia; Email: babin@iae.nsk.su. 

\end{document}